\documentclass[aps, prl, twocolumn, superscriptaddress, letterpaper]{revtex4-2}

\usepackage{graphicx}
\usepackage{amssymb}
\usepackage{amsmath}
\usepackage{float}
\usepackage{txfonts}
\usepackage{bm}
\usepackage{color}
\usepackage[colorlinks=true, linkcolor=blue, anchorcolor=black, citecolor=blue, filecolor=black, menucolor=black, runcolor=black, urlcolor=blue]{hyperref}
\usepackage{mathrsfs}

\begin{document}

\title{Observation of $\pi/2$ modes in an acoustic Floquet system}

\author{Zheyu Cheng}
\thanks{These authors contribute equally.}
\affiliation{Division of Physics and Applied Physics, School of Physical and Mathematical Sciences, Nanyang Technological University,
Singapore 637371, Singapore}

\author{Raditya Weda Bomantara}
\thanks{These authors contribute equally.}
\affiliation{Centre for Engineered Quantum Systems, School of Physics, University of Sydney, Sydney, New South Wales 2006, Australia}

\author{Haoran Xue}
\affiliation{Division of Physics and Applied Physics, School of Physical and Mathematical Sciences, Nanyang Technological University,
Singapore 637371, Singapore}

\author{Weiwei Zhu}
\affiliation{Department of Physics, National University of Singapore, Singapore 117542, Singapore}

\author{Jiangbin Gong}
\email{phygj@nus.edu.sg}
\affiliation{Department of Physics, National University of Singapore, Singapore 117542, Singapore}
\affiliation{Centre for Quantum Technologies, National University of Singapore, 117543, Singapore}

\author{Baile Zhang}
\email{blzhang@ntu.edu.sg}
\affiliation{Division of Physics and Applied Physics, School of Physical and Mathematical Sciences, Nanyang Technological University,
Singapore 637371, Singapore}
\affiliation{Centre for Disruptive Photonic Technologies, Nanyang Technological University, Singapore 637371, Singapore}

\begin{abstract}
Topological phases of matter have remained an active area of research in the last few decades. Periodic driving is known to be a powerful tool for enriching such exotic phases, which leads to various phenomena with no static analogs. One such phenomenon is the emergence of the elusive $\pi/2$ modes, i.e., a type of topological boundary state pinned at a quarter of the driving frequency. The latter may lead to the formation of Floquet parafermions in the presence of interaction, which is known to support more computational power than Majorana particles. In this work, we experimentally verify the signature of $\pi/2$ modes in an acoustic waveguide array, which is designed to simulate a square-root periodically driven Su-Schrieffer-Heeger model. This is accomplished by confirming the $4T$-periodicity ($T$ being the driving period) profile of an initial-boundary excitation, which we also show theoretically to be the smoking gun evidence of $\pi/2$ modes. Our findings are expected to motivate further studies of $\pi/2$ modes in quantum systems for potential technological applications.  
\end{abstract}

\maketitle

\emph{Introduction.} Topology plays a significant role in condensed matter physics through its ability to protect certain physical properties against perturbations. Its presence has been identified in a variety of systems including topological insulators \cite{bernevig2006quantum, qi2008topological}, topological semimetals \cite{burkov2011weyl, wan2011topological}, and topological superconductors \cite{hasan2010colloquium, qi2011topological}. These topological phases share one defining feature, i.e., they support robust boundary modes protected by a global quantity, i.e., the topological invariant. 

Since the last decade, the implementation of periodic driving in studies of topological phases has been extensively pursued \cite{oka2009photovoltaic, kitagawa2010topological, lindner2011floquet, ho2012quantized, kitagawa2012observation, rechtsman2013photonic, jotzu2014experimental, ho2014topological, bomantara2016generateing, cheng2019observation, long2019floquet, mciver2020light, rudner2013anomalous, nathan2015topological,zhu2021, bomantara2021z4, zhu2022time,po2016chiral, jiang2011majorana}. Not only does periodic driving have the capability to turn an otherwise normal system into a topological one \cite{oka2009photovoltaic, kitagawa2010topological,  lindner2011floquet}, but it can also generate unique topological features with no static counterparts \cite{jiang2011majorana, rudner2013anomalous, nathan2015topological, po2016chiral, bomantara2020time,zhu2021, bomantara2021z4, zhu2022time}. The latter is made possible by the periodicity nature of quasienergy, i.e., the analog of energy in time-periodic systems. It prominently includes the coexistence of boundary zero and $\pi$ modes which are respectively pinned at quasienergy of zero and a half the driving frequency. Such a feature has recently been exploited for quantum computing applications \cite{bomantara2018simulation, bomantara2018quanum, bomantara2020combating, bomantara2020measurement}. 

Periodically driven (hereafter referred to as Floquet) systems can support even richer topological features beyond the above zero and $\pi$ modes. Of particular interest are the so-called $\pi/2$ modes, i.e., boundary modes which are pinned at a quarter of the driving frequency. While the more common zero and $\pi$ modes may only generate Majorana particles in some Floquet systems, $\pi/2$ modes underlie the formation of the more exotic parafermion edge modes in the presence of interaction \cite{bomantara2021z4, sreejith2016parafermion}. Given that parafermion modes are known to support richer topologically protected quantum gate operations than their Majorana counterparts \cite{hutter2016quantum}, the experimental realization of $\pi/2$ modes thus serves as a first step towards developing a better topological quantum computer. However, in previous studies, $\pi/2$ modes were proposed either in a very elaborate driven system \cite{bomantara2021z4} or in a strongly interacting system \cite{sreejith2016parafermion}, hindering their accessibility in experiments.

In this work, we utilize the square-root procedure of Ref.~\cite{bomantara2021square} to propose and experimentally realize a simple Floquet system in an acoustic waveguide array capable of supporting the elusive $\pi/2$ modes, along with zero and $\pi$ modes simultaneously. In the presence of these modes, we theoretically show and experimentally verify that a state injected at one end of the system displays a $4T$-periodicity. Our experiments not only confirm the existence of $\pi/2$ modes but also capture their interplay with existing zero and $\pi$ modes.

\begin{figure}[b]
\centering
\includegraphics[width = 0.48 \textwidth]{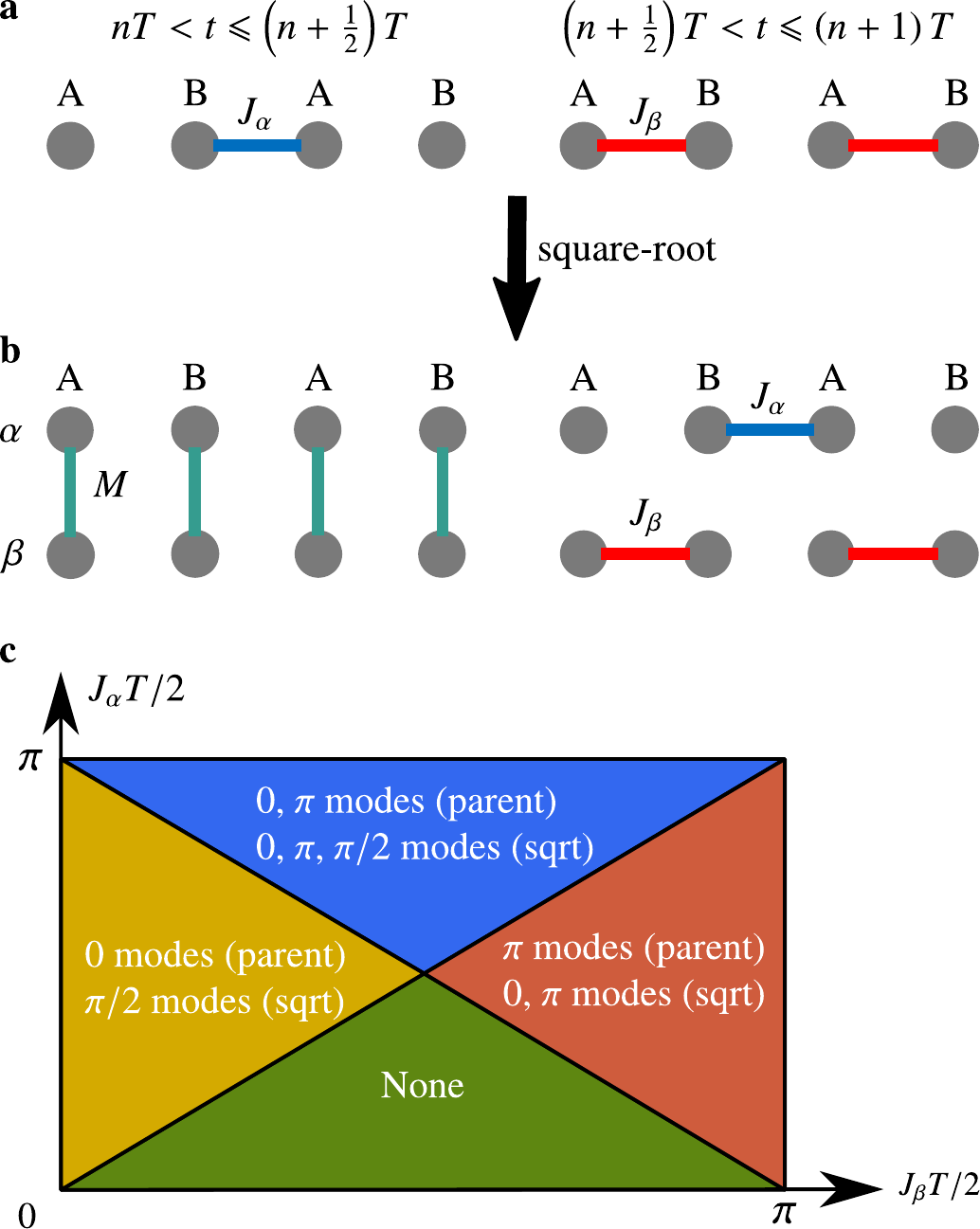}%
\caption{\textbf{a,b}. Schematic of (a) Floquet SSH model and (b) its square-root counterpart. \textbf{c}. Phase diagram for square root and parent Floquet system.}
\label{fig01}
\end{figure}

\emph{Square-root Floquet SSH model.} We consider a system of two one-dimensional (1D) chains subjected to the two-time-step Hamiltonian 
\begin{equation}
{H^\text{root}}\left( t \right) = \left\{ {\begin{array}{*{20}{l}}
  h_1^\text{root} &\qquad nT < t \leqslant \left(n + \frac{1}{2}\right)T \\ 
  h_2^\text{root} &\qquad \left(n + \frac{1}{2}\right)T < t \leqslant \left( {n + 1} \right)T 
\end{array}} \right.,
\label{01}
\end{equation}
where
\begin{align}
  h_1^{{\text{root}}} =& \sum_{j=1}^N \sum_{S=A,B} M|j,S,\alpha\rangle \langle j,S,\beta | +h.c.  , \label{02} \\
  h_2^{{\text{root}}} =& \sum_{j=1}^{N-1} J_\alpha  |j,B,\alpha\rangle \langle j+1,A,\alpha | + \sum_{j=1}^{N} J_\beta |j,A,\beta \rangle \langle j,B,\beta | +h.c.,  \label{03}
\end{align}
$|j,S,\xi\rangle$ denotes a state at sublattice $S=A,B$ of the $j$th site in the chain species $\xi=\alpha,\beta$, and $n \in \mathbb{Z}$. It is schematically shown in Fig.~\ref{fig01}b and can be understood as a nontrivial square-root of a Floquet Su-Schrieffer-Heeger (SSH) model (see Fig.~\ref{fig01}a). Specifically, the parent model describes a system of \emph{a single} 1D chain subjected to a two-time-step Hamiltonian of the form Eq.~(\ref{01}), such that during the first- and second-half of the period, it is respectively given by
\begin{align}
  h_1^\text{parent} =& {J_\alpha }|j,B\rangle \langle j+1,A | +h.c.\;, \label{04}\\
  h_2^\text{parent} =& {J_\beta }|j,A\rangle \langle j,B | +h.c.\;. \label{05} 
\end{align}
The associated one-period time-evolution operator (hereafter referred to as the Floquet operator) is then given by 
\begin{equation}
    U^{\rm parent} = U_2^{\rm parent} U_1^{\rm parent}\;, \;\text{where } U_j^{\rm parent}=\exp\left(-\mathrm{i} h_j^{\rm parent}\frac{T}{2}\right) \;.
\end{equation}
 The use of an additional chain in our square-root model then introduces an ancillary degree of freedom that facilitates the square-rooting procedure \cite{bomantara2021square} in the spirit of obtaining Dirac equation \cite{dirac1928quantum} from Klein-Gordon equation \cite{klein1927elektrodynamik,gordon1926comptoneffekt} (see also Ref.~\cite{arkinstall2017topological}, as well as related theoretical \cite{ezawa2020systematic,pelegri2019topological, ding2019recent, ezawa2020systematic, mizoguchi2020square, marques2021one, lin2021square, marques20212n, yoshida2021square, mizoguchi2021square, dias2021matryoshka} and experimental \cite{kremer2020square, yan2020acoustic} studies). 
 
 Eq.~(\ref{01}) can be intuitively understood as follows. A particle initially living in chain $\alpha$ ($\beta$) evolves under $h_1^\text{parent}$ ($h_2^\text{parent}$) for the first half of the period and hops to the other chain in the second half of the period. The particle, which is now in chain $\beta$ ($\alpha$), then evolves under $h_2^\text{parent}$ ($h_1^\text{parent}$) for another half period and hops back to the original chain $\alpha$ ($\beta$) at the end of the second period. Therefore, when viewed over two periods, the particle effectively evolves one full period under the parent Hamiltonian. On the other hand, over one period, the particle undergoes a generally nontrivial evolution that can give rise to new physics, including $\pi/2$ modes, the main focus of this work.
 
Mathematically, at $MT = (2m+1)\pi$ with $m\in \mathbb{Z}$, the Floquet operator associated with Eq.~(\ref{01}) can be easily obtained as
\begin{equation}
{U^\text{root}} = \exp \left[ { - ih_2^\text{root}\frac{T}{2}} \right]\exp \left[ { - ih_1^\text{root}\frac{T}{2}} \right] = \left( {\begin{array}{*{20}{c}}
  0&{ - iU_1^\text{parent}} \\ 
  { - iU_2^\text{parent}}&0 
  \end{array}} \right)
\label{06}
\end{equation}
Indeed, up to a unitary transformation, $U^{\rm parent}$ is reproduced by ${\left(U^\text{root}\right)}^2 = -\text{diag}\left( {U_1^\text{parent}U_2^\text{parent},U_2^\text{parent}U_1^\text{parent}} \right)$, as expected from our intuition above. As $MT$ deviates from $(2m+1)\pi$, ${\left(U^\text{root}\right)}^2$ is no longer diagonal and directly related to $U^\text{parent}$. However, owing to the robustness of Floquet phases and as numerically demonstrated in Ref.~\cite{bomantara2021square}, ${\left(U^\text{root}\right)}^2$ still exhibits the physics expected from its parent system provided $MT-(2m+1)\pi$ is not too large. 

Depending on the system parameter values, the parent Floquet SSH model may support either a pair of zero modes only, a pair of $\pi$ modes only, a pair of zero modes and a pair of $\pi$ modes, or neither of zero modes nor $\pi$ modes. As elucidated in Ref.~\cite{bomantara2021square}, the presence of $\pi$ modes in the parent system leads to the coexistence of zero modes and $\pi$ modes in its square-root counterpart, whereas the presence of zero modes in the former leads to the emergence of the elusive $\pi/2$ modes in the latter. Consequently, the proposed square-root Floquet SSH model possesses the phase diagram shown in Fig.~\ref{fig01}c (see Supplemental Material (SM) for its derivation).  

\begin{figure*}
\centering
\includegraphics[width =  \textwidth]{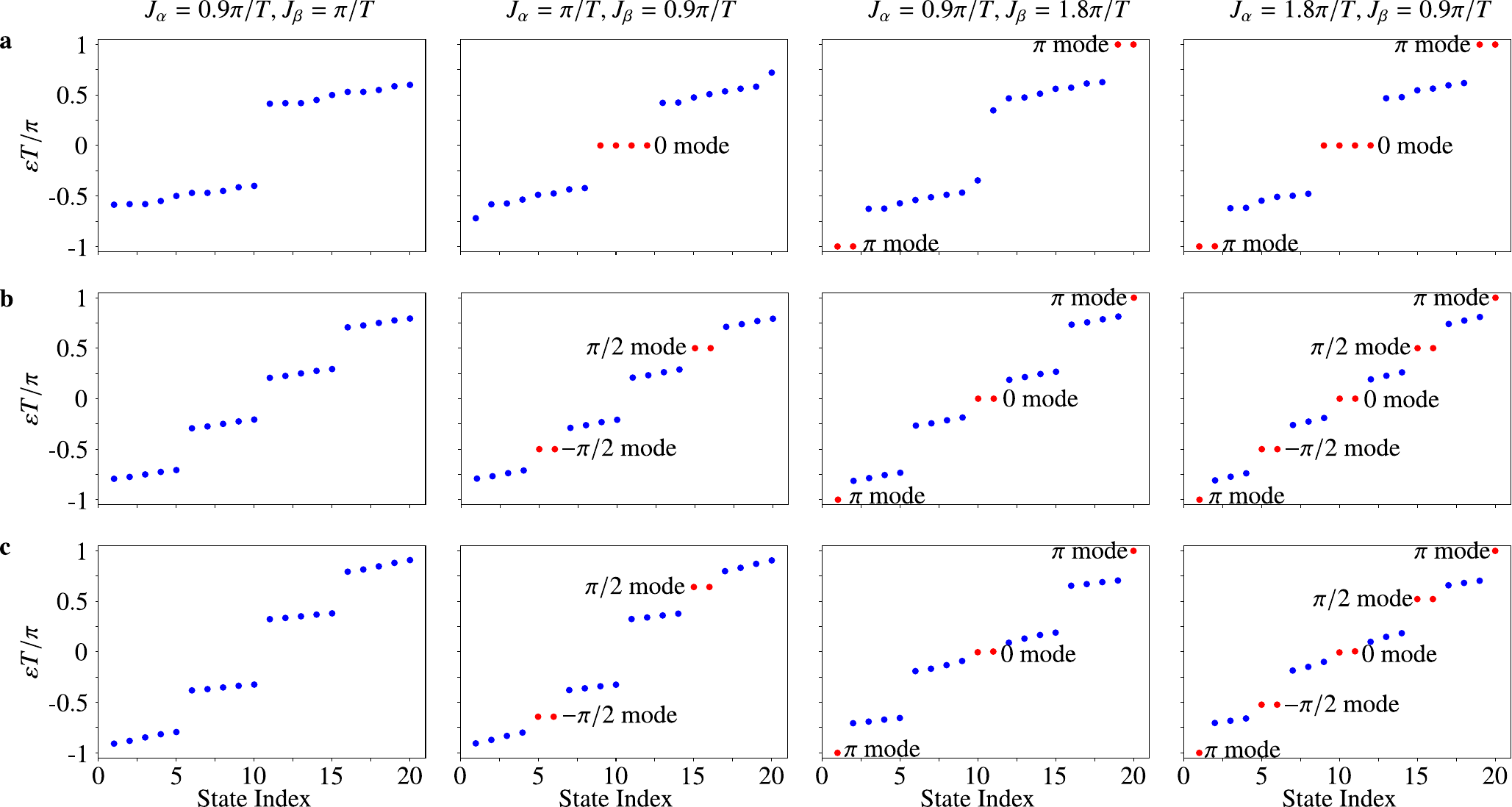}%
\caption{\textbf{a}. Illustration of quasi-energy spectrum for two decoupled copies of original Floquet system, each containing 10 sites. From left to right: $J_\alpha= 0.9\pi/T, J_\beta = \pi/T$; $J_\alpha= \pi/T, J_\beta = 0.9\pi/T$; $J_\alpha= 0.9\pi/T, J_\beta = 1.8\pi/T$; $J_\alpha= 1.8\pi/T, J_\beta = 0.9\pi/T$.  \textbf{b,c}. Illustration of quasi-energy spectrum for square-root Floquet system with 20 sites under (b) $M = \pi/T$, (c) $M = 1.3\pi/T$, and the same parameters $J_\alpha$ and $J_\beta$ as panel \textbf{a}.}
\label{fig02}
\end{figure*}

\emph{Topological edge modes.} In Fig.~\ref{fig02}(a-c), we numerically calculate the quasienergy spectrum of the Floquet SSH model and its square-root counterpart at a representative point from each of the four phases in Fig.~\ref{fig01}c. There, in-gap quasienergy solutions at $0$, $\pm \pi/T$, and $\pm \pi/(2T)$ correspond to zero, $\pi$, and $\pi/2$ modes respectively, whose edge-localization nature is confirmed in Fig.~\ref{fig02}(b-c) and SM. Fig.~\ref{fig02}(c) further verifies the robustness of zero, $\pi$, and $\pi/2$ modes in the square-root model against imperfection in the parameter $M$. There, all modes remain gapped from the rest of the bulk spectrum, with zero and $\pi$ modes additionally remaining exactly pinned $0$ and $\pi/T$ quasienergy respectively, whereas $\pi/2$ modes are slightly shifted away from $\pm \pi/(2T)$ quasienergy. 

The quasienergy rigidity of the system's zero and $\pi$ modes originates from the presence of chiral symmetry inherited from its parent model. Indeed, by writing the system's Floquet operator in the momentum space and under the symmetric time frame \cite{asboth2012symmetries, asboth2013bulk}, i.e., the shift of time origin to $t_0=T/4$, we obtain
\begin{equation}
    \mathcal{U}^{\rm root}(k) = e^{ - \mathrm{i} \mathcal{H}_1^\text{root}(k)\frac{T}{4}}  e^{ - \mathrm{i} \mathcal{H}_2^\text{root}(k)\frac{T}{2}}e^{ - \mathrm{i} \mathcal{H}_1^\text{root}(k)\frac{T}{4}} \;,
\end{equation}
where 
\begin{eqnarray}
    \mathcal{H}_2^\text{root}(k) &=& {J_\alpha }\frac{{{\tau _0} + {\tau _3}}}{2}  \left( {\cos k{\sigma _1} + \sin k{\sigma _2}} \right) + {J_\beta }\frac{{{\tau _0} - {\tau _3}}}{2}  {\sigma _1} \;, \nonumber \\ 
    \mathcal{H}_1^\text{root}(k) &=& M\tau_1 \;, 
\end{eqnarray}
 $\tau_{1/2/3}$ and $\sigma_{1/2/3}$ are Pauli matrices acting on the chain species and sublattice subspace respectively, and $k$ is the quasimomentum. It can then be verified that $\mathcal{C}\mathcal{U}^{\rm root}(k) \mathcal{C}^\dagger = \left(\mathcal{U}^{\rm root}(k)\right)^\dagger$, where $\mathcal{C}= \sigma_3 \tau_3$ is the chiral symmetry operator. As a result, the system's quasinergies come in pairs of $\varepsilon$ and $-\varepsilon$. Moreover, the special values $\varepsilon=0$ and $\varepsilon=\pi/T$ are at least two-fold degenerate and can be chosen to be simultaneous $\pm 1$ eigenvalues of $\mathcal{C}$. The discreteness of $\mathcal{C}$ eigenvalues in turn protects the resulting degenerate eigenstates, i.e., zero and $\pi$ modes respectively. 

At $MT=(2m+1)\pi$, there exists an additional subchiral symmetry that operates as $\mathcal{C}_{(1/2)} \mathcal{U}^{\rm root}(k) \mathcal{C}_{(1/2)}^\dagger = -\mathcal{U}^{\rm root}(k)$, where $\mathcal{C}_{(1/2)}=\tau_3$ \cite{marques2019one, zhou2022q}. It guarantees that the quasienergies of $\mathcal{U}^{\rm root}(k)$ come in pairs of $\varepsilon$ and $\varepsilon\pm \pi/T$. Consequently, the special values $\varepsilon = \pm \pi/(2T)$ are at least two-fold degenerate and can be chosen as simultaneous $\pm 1$ eigenvalues of $\mathcal{C}\cdot\mathcal{C}_{(1/2)}$. The discreteness of $\mathcal{C}\cdot\mathcal{C}_{(1/2)}$ eigenvalues then protects the resulting $\pi/2$ modes. Such a subchiral symmetry no longer exists at $MT\neq (2m+1)\pi$, thus allowing $\pi/2$ modes to generally deviate from their expected $\pm \pi/(2T)$ quasienergy values. However, as demonstrated in the SM, at small $\epsilon \equiv |MT-(2m+1)\pi|$, the shift in the quasienergy of the $\pi/2$ modes in the blue regime of Fig.~\ref{fig01} is at least second order in $\epsilon$. This, coupled with the fact that their eigenmodes profile is similar to the ideal $MT=(2m+1)\pi$ case, justifies the term $\pi/2$ modes even as $MT=(2m+1)\pi$ condition cannot be exactly achieved in our experiments.

The coexistence of zero, $\pm \pi/2$, and $\pi$ modes, i.e., in the blue regime of Fig.~\ref{fig01}(c), leads to a dynamical signature that we manage to observe in our experiment. In particular, by noting that $U^{\rm root} |0\rangle =|0\rangle$, $U^{\rm root} |\pm \pi/2\rangle =\mp \mathrm{i} |\pm \pi/2\rangle$, $U^{\rm root} |\pi\rangle =-|\pi\rangle$, for zero, $\pm \pi/2$, and $\pi$ modes respectively, any superposition $|\psi\rangle = a |0\rangle + b |\pi/2\rangle + c |\pi\rangle +d |-\pi/2\rangle$ yields $4T$-periodicity. Indeed, over the course of period, each component of $|\psi\rangle$ acquires a relative phase of $\pi/2$, thus transforming it into a distinct state. After four periods, such a relative phase accumulates into $2\pi$, thus recovering the state $|\psi \rangle$. On the other hand, in the absence of $\pm \pi/2$ modes, i.e., the red regime of Fig.~\ref{fig01}(c), the remaining coexistence of zero and $\pi$ modes can similarly be probed by the $2T$-periodicity signature of a state comprising a superposition of $|0\rangle$ and $|\pi\rangle$.      

As mathematically detailed in the SM, the state $|1,A,\alpha\rangle$ is exactly $\propto |0\rangle + |\pi/2\rangle + |\pi\rangle - |\pi/2\rangle$ or $\propto |0\rangle + |\pi\rangle $ at special parameter values within the blue or red regime of Fig.~\ref{fig01}(c) respectively. Away from these special parameter values, the localized nature of edge modes implies that $|1,A, \alpha\rangle$ still has considerable support on the respective superposition state above, thus preserving the expected $4T$-periodicity and $2T$-periodicity throughout most of the blue and red regime in Fig.~\ref{fig01}(c) respectively. This argument is further supported by our numerics in SM, which directly evolves $|1,A,\alpha\rangle$ under Eq.~(\ref{01}) at two general parameter values within the blue and red regime of Fig.~\ref{fig01}(c) respectively. 

\begin{figure*}
\centering
\includegraphics[width =  \textwidth]{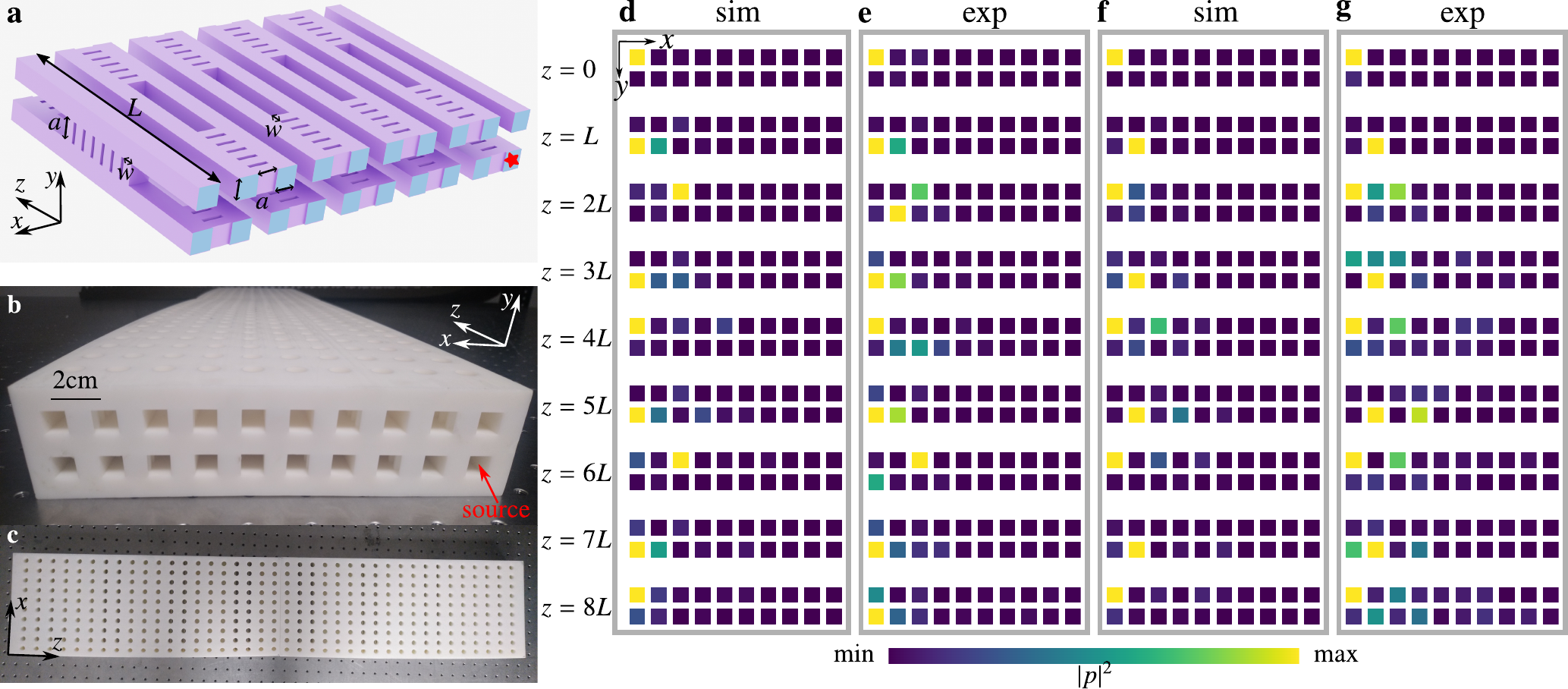}%
\caption{\textbf{a}. One Floquet period in the waveguide array, which hosts $0,\pi/2, \pi$ modes. An initial excitation was put at the red star in the simulation. The purple color is the hard boundary, and the light blue represents the hollow region filled with air. \textbf{b, c}. Photo of the fabricated sample with plugs removed. The source was put at the lower-right corner in the measurement.  \textbf{d, f}. Mode evolution of fullwave simulation with $0,\pi/2, \pi \; (0,\pi)$ modes appear at 8kHz accordingly. \textbf{e, g}. Experimental counterparts of panel (d) and (f) respectively. The effective parameter values for the simulation and experimental results are \textbf{(d,e)} $J_\alpha=1.668\pi/T$, $J_\beta=0.850\pi/T$, $M=0.976\pi/T$ and \textbf{(f,g)}, $J_\alpha=0.850\pi/T$, $J_\beta=1.668\pi/T, M=0.976\pi/T$. }
\label{fig03}
\end{figure*}

\emph{Experimental realization in acoustic waveguides.} We will now present our experimental observation of $\pi/2$ modes in an acoustic waveguide array. To facilitate the theoretical description of our experiment, we consider the slowly varying amplitude (SVA) approximation $\left| {\frac{{{\partial ^2}p}}{{\partial {z^2}}}} \right| \ll {k_z}\left| {\frac{{\partial p}}{{\partial z}}} \right|$ and take $p \to pe^{ik_z z}$ to the Helmholtz equation $\left( {{\nabla ^2} + {{\bf{k}}^2}} \right)p = 0$, where $p$ is the acoustic pressure. After SVA approximation, we get the paraxial wave equation, which can be written in the Schr{\"o}dinger form
\begin{equation}
i\frac{{\partial p}}{{\partial z}} = {H_\text{eff}}p = \left( { - \frac{{\nabla _ \bot ^2}}{{2{k_z}}} - \frac{{{{\bf{k}}^2} - k_z^2}}{{2{k_z}}}} \right)p,
\label{07}
\end{equation}
with $\nabla _ \bot ^2 = \partial _x^2 + \partial _y^2$ as the 2D Laplacian operator. By further employing a tight-binding approximation, Eq.~(\ref{07}) can then simulate the lattice model of Eq.~(\ref{01}) with tunable hopping amplitudes (see SM). Here, ``time'' is simulated by ``propagation direction'' $z$. This scheme is the ``paraxial acoustics'', which is analogous to the ``paraxial photonics'' \cite{jani2000paraxial, miguel2009paraxial, carnicer2012on, chane2021demonstrating}. By periodically modulating the shape of the waveguides along the $z$ direction, a Floquet system can be simulated. In this case, the wave function dynamics can be probed by detecting the pressure in the $z$ direction.

To realize the acoustic analogy of Eq.~(\ref{01}), we construct the acoustic waveguide array (or 2D resonator system) as shown in Figs.~\ref{fig03}(a-c). The whole structure is covered by hard walls and filled with the air of density $1.8\rm{kg/m^3}$ and sound velocity $347\text{m/s}$. The length of one Floquet period is $L=133\text{mm}$, and $8$ Floquet periods are used in the simulation and experiment. In this setup, the waveguides (2D resonators) and link tubes simulate the lattice sites and nearest-neighboring hopping respectively. As detailed in the SM, the geometry of waveguides and the number of link tubes between two adjacent waveguides determine the hopping strength ($J_\alpha, J_\beta, M$ in Eq.~(\ref{01})), respectively. Here, for example, $J_\alpha = 0.850\pi/T$ can be satisfied by $6$ link tubes at 8kHz, where $T$ is the Floquet period in acoustic structure. The cross profile of the waveguide is a square with a side length $a=10$mm. The height and width of the coupling tubes are $a$ and the length is $w=5$mm (see Fig.~\ref{fig03}a). After this design, the resonant frequency is 8kHz in our acoustic waveguide array.

We fabricated two experimental samples with different effective parameter values via 3D printing technology. For each sample, we drill $780$ identical and equally spaced holes, inside which a microphone is inserted and responds to pressure (see Fig.~\ref{fig03}c).  When not in use, these holes are covered by plugs. To numerically verify the dynamical signature of the $\pi/2$ edge modes, a speaker is placed at $z = 0$ on the lower-right corner (indicated by a red arrow in Fig.~\ref{fig03}b), which generates an initial state $|1,A,\alpha\rangle$. By measuring the acoustic pressure on the sample at some specific $z$ values, the stroboscopic time evolution profile of the effective square-root Floquet SSH model can be probed. Our experimental results are summarized in Figs.~\ref{fig03}(e) and (g), which are according to the blue and red regime of Fig.~\ref{fig01}(c), respectively. Consistent with our full-wave simulation (Fig.~\ref{fig03}(d,f)) , the presence and absence of $\pi/2$ modes manifest themselves in $4T$-and $2T$-period behavior of Figs.~\ref{fig03}(e) and (g) respectively.    

\emph{Concluding remarks.} In this paper, we have proposed and experimentally realized a square-root Floquet SSH model, which exhibits `$\pi/(2T)$ quasienergy edge states' termed the $\pi/2$ modes, in coexistence with the more common zero and $\pi$ modes. These exotic $\pi/2$ modes originate from the zero modes in the parent system and can thus be predicted by the topological invariant characterizing the latter, i.e., $\nu_0$ as defined in the SM. We have further identified a subchiral symmetry that protects these $\pi/2$ modes, as well as their dynamical signature, both of which enable their successful observation in our acoustic experiment. 

Recently, Floquet topological systems which support coexisting zero and $\pi$ modes are shown to be advantageous for quantum computation \cite{bomantara2018quanum, bomantara2020measurement, bomantara2020combating}. Moreover, the experimentally verified $\pi/2$ modes have been identified as the building blocks of $Z_4$ parafermions \cite{bomantara2021z4}, which support a richer set of topologically protected gates as compared with Majorana particles. The realization of coexisting zero, $\pi/2$, and $\pi$ modes reported in this paper thus opens up exciting opportunities for superior topological quantum computing beyond Majorana particles. While the present experiment concerns a classical system, it is expected that some of our findings can be carried over to the quantum realm. To this end, the simulation of braiding among the various edge modes we observed in our acoustic setup, e.g., achieved by adapting the technique of Ref.~\cite{chen2022classical}, may present a promising future direction. 

\begin{acknowledgments}
	{\bf Acknowledgement}: Z.C., H.X., and B.Z. are supported by the Singapore Ministry of Education Academic Research Fund Tier 2 Grant MOE2019-T2-2-085, and Singapore National Research Foundation Competitive Research Program Grant NRF-CRP23-2019-0007. R.W.B is supported by the Australian Research Council Centre of Excellence for Engineered Quantum Systems (EQUS, CE170100009).  W.Z. and J.G. are funded by the Singapore National Research Foundation Grant No. NRF-NRFI2017-04 (WBS No. R-144-000-378- 281).
\end{acknowledgments}

\bibliography{manuscript}

\end{document}